\documentclass[aps,preprint,showpacs,superscriptaddress,groupedaddress]
{elsarticle}
\usepackage{bm}       
\usepackage{amssymb}
\usepackage{amsmath}     
\usepackage{hyperref} 
\usepackage{psfrag}
\usepackage[cp1250]{inputenc}
\usepackage{graphicx}
\usepackage{color}
\DeclareMathAlphabet{\mathpzc}{OT1}{pzc}{m}{it}

\begin{document}
\title{\boldmath Endlessly flat scalar potentials and $\alpha$-attractors}

\author[cracow]{Micha{\l} Artymowski} 
\author[heidelberg]{Javier Rubio} 

\address[cracow]{Institute of Physics, Jagiellonian University, {\L}ojasiewicza 11, 30-348 Krak{\'o}w, Poland}
\address[heidelberg]{Institut f\"ur Theoretische Physik, Ruprecht-Karls-Universit\"at Heidelberg\\
Philosophenweg 16, 69120 Heidelberg, Germany.}

\begin{abstract}
We consider a minimally-coupled inflationary theory with a general 
scalar potential $V(f(\varphi))= V(\xi\sum_{k=1}^{n}\lambda_k \varphi^k)$ containing 
a stationary point of maximal order $m$. We show that asymptotically flat potentials can be associated 
to stationary points of infinite order and discuss the relation of our approach to the theory of $\alpha$-attractors.
\end{abstract}

\maketitle

\section{Introduction} \label{sec:Introduction}
Cosmic inflation \cite{Starobinsky:1980te,Mukhanov:1981xt,Guth:1980zm,Linde:1981mu,Albrecht:1982wi,Linde:1983gd}
is nowadays a well established paradigm able to solve most of the hot Big Bang puzzles and to 
explain the generation of the almost scale invariant spectrum of coherent primordial perturbations giving 
rise to structure formation  \cite{Ade:2015lrj} (for a review, see for instance~\cite{Lyth:1998xn,Mazumdar:2010sa}). 

The vast majority of inflationary models assume the early domination 
of a scalar field $\varphi$ with a sufficiently flat potential $V(\varphi)$. For canonically normalized fields, the 
inflationary observables can be parametrized in terms of the so-called slow-roll parameters
\begin{equation}\label{inf_cond}
\epsilon = \frac{1}{2}\left(\frac{V'}{V}\right)^2 \,, \qquad\qquad \eta = \frac{V''}{V} \, ,
\end{equation}
where the primes denote derivatives with respect to the inflaton field $\varphi$. For inflation to 
take place, the \textit{slow-roll}  conditions $\epsilon, |\eta| \ll 1$  must be satisfied. Note 
that these requirements should be understood as \textit{local} conditions on an arbitrary potential, which 
can in principle contain a large number of extrema and slopes. 

Locally flat regions in the potential appear generically in the vicinity of stationary points. Even though 
saddle-point models of inflation have been shown to be inconsistent with the data \cite{Allahverdi:2006we}, 
one should not exclude the appearance of higher order points on the inflationary potential~\cite{Hamada:2015wea}.

In this paper we will take a model building perspective. Rather than asking about the origin of  $V(\varphi)$, we will require
it to \textit{locally} satisfy some particular flatness conditions. We will ask for the existence of a 
single stationary point without imposing any further restrictions on the shape of the potential. Similar 
studies have been performed in the context of modifed gravity theories. In particular, it was shown in 
Refs.~\cite{Artymowski:2015pna,Artymowski:2016mlh} that the requirement of vanishing derivatives in a $f(R)$ model gives rise 
to an inflationary plateau  in the Einstein frame formulation of the theory. The main purpose of this 
short letter is to extend this analysis to the scalar sector and to make explicit the equivalence between the 
stationary point picture and the $\alpha$-attractor formulation~\cite{Carrasco:2015rva,Kallosh:2014rga,Carrasco:2015pla,Kallosh:2015lwa,Linde:2015uga}.

The structure of this paper is as follows. In Sec.~\ref{sec:general} we construct a general scalar inflationary 
theory containing a stationary point of a given order $m$. The equivalence of these theories and the  
$\alpha$-attractor formulation is presented in Sec.~\ref{sec:alpha}. Finally, we summarize in Sec.~\ref{sec:Summary}.

%%%%%%%%%%%%%%%%%%%%%%%%%%%%%%%%%%%%%%%%%%%%%%%%%%
%%%%%%%%%%%%%%%%%%%%%%%%%%%%%%%%%%%%%%%%%%%%%%%%%%

\section{Stationary point inflation for general form of scalar potential} \label{sec:general}

Let us assume the early Universe to be approximately described by standard Einstein gravity and a
homogeneous inflaton field $\varphi$. The associated Lagrangian density in Planckian units reads\footnote{We use the 
convention $8\pi G = M_{P}^{-2} = 1$ with $M_{P} = 2.435\times 10^{18}$ GeV the reduced Planck mass.}
\begin{equation}\label{eq:lagrang}
\frac{\cal L}{\sqrt{-g}} =\frac{1}{2}R + \frac{1}{2}(\partial\varphi)^2-V(f(\varphi)) \, ,
\end{equation}
with
\begin{equation}
f(\varphi) = \xi \sum_{k=1}^n \, \lambda_k \, \varphi^k \, ,\label{eq:lambdak}
\end{equation}
and $\lambda_k$ constants.\footnote{Note that one can always redefine the $\xi$ constant such that $\lambda_1 = 1$.}
The upper-limit in the power-law expansion \eqref{eq:lambdak} implicitly assumes the existence of 
some symmetry or hierarchy of scales effectively suppressing terms with $k > n$. Note however that this constraint 
does not restrict the maximal power of $\varphi$ appearing in the scalar potential $V(\varphi)$, which can in 
principle contain higher powers of the scalar field.\footnote{For instance, if the potential is chosen 
as $V \propto f^m$, the highest power of $\varphi$ is $n\times m$.}

Consider an inflationary potential $V(f(\varphi))$ containing a stationary point of order $m$ at some field value 
$\varphi = \varphi_s$. This stationary  point could  be a local maximum or a saddle point able to give rise to inflation in the flat area surrounding it. As long as all derivatives of the potential with respect to $f$ are well-defined, the existence of a higher order 
stationary point in $V(f(\varphi))$ translates into a set of conditions on $f$
\begin{equation}\label{eq:Vconst}
\frac{d^mV}{d\varphi^m} = 0\quad \Leftrightarrow \quad \frac{d^m f}{d\varphi^m} = 0\,.
\end{equation}

For any integer $n$ in Eq.~\eqref{eq:lambdak}, Eq.~\eqref{eq:Vconst} provides at most $m=n-1$ constraints. The minimal 
scenario is the so-called saddle-point inflation for which $V' = V'' = 0$ at $\varphi=\varphi_s$. Unfortunately,  the 
small value of the spectral tilt predicted by this model is inconsistent with the data~\cite{Allahverdi:2006we}. Note however, 
that a $m$-order stationary point may be still a viable source of inflation \cite{Hamada:2015wea}. Demanding the existence of a $m=n-1$ stationary point leads to the following form for $\varphi_s$ and the $\lambda_k$ 
coefficients
\begin{equation}
\varphi_s =\left(n \, \lambda\right)^{\frac{-1}{n-1}}\, , 
\qquad \lambda_k = (-1)^{k+1}\frac{(n-1)!}{k!(n-k)!}(n \, \lambda)^{\frac{k-1}{n-1}} \, , \label{eq:lambdaksaddle}
\end{equation}
with $\lambda\equiv\lambda_n$ a free parameter to be fixed by observations. Inserting this result into 
Eq.~(\ref{eq:lambdak}) we get the following expression for $f(\varphi)$,
\begin{equation}
f(\varphi) = \frac{\xi}{n}\left(n \, \lambda\right)^{\frac{-1}{n-1}}\left(1 - \left(1 -
\left(n \, \lambda\right)^{\frac{1}{n-1}}\varphi \right)^n\right) \,, \label{eq:fgeneral}
\end{equation}
which can be redefined by an additive constant without any loss of generality.

In the most general case the coefficients $\lambda_k$ in Eq.~\eqref{eq:lambdak} can depend on $n$ and $\xi$. This property may influence not only the
primordial inhomogeneities but also the convergence of $f$ in the $n \to \infty$ limit and/or the perturbativity of the 
theory. We will restrict ourselves to two forms on $\lambda$, namely to $\lambda =\lambda_1 \equiv1/n$ 
and $\lambda = \lambda_2 \equiv \frac{1}{\xi}(\xi/n)^{n}$. For these two cases, 
Eq.~(\ref{eq:fgeneral}) simplifies to\footnote{The structure of the $\lambda_k$ coefficients  giving rise to Eqs.~\eqref{eq:phin} and \eqref{eq:fn^-n} may seem \textit{ad-hoc} from the bottom-up perspective considered in this paper. However, this type of structure could arise naturally from a fundamental theory able to generate the appropriate potential after integrating out irrelevant degrees of freedom. A particular example is the Starobinsky-like model of inflation $V(\varphi) = M^2\left(1-e^{-\xi \varphi}\right)^2$, where all the coefficients $\lambda_k$ are related to a single parameter $\xi$.}

\begin{eqnarray}
f(\varphi) = \frac{\xi}{n}\left(1 - \left(1 - \varphi \right)^n\right) \, ,\qquad \varphi_s = 
1\qquad &\text{for}& \qquad \lambda = \lambda_1 \, , \label{eq:phin} \\
f(\varphi) =1- \left(1 - \frac{\xi}{n}\varphi \right)^n \, ,\qquad \varphi_s = 
\frac{n}{\xi} \qquad &\text{for}& \qquad \lambda = \lambda_2 \, . \label{eq:fn^-n}
\end{eqnarray}
Note that  for $\xi = n$, $\xi > n$ and $\xi<n$ one obtains respectively $\lambda_1 = \lambda_2$, 
$\lambda_1 < \lambda_2$ and $\lambda_1 > \lambda_2$.

\begin{figure}
\centering
\includegraphics[height=4.6cm]{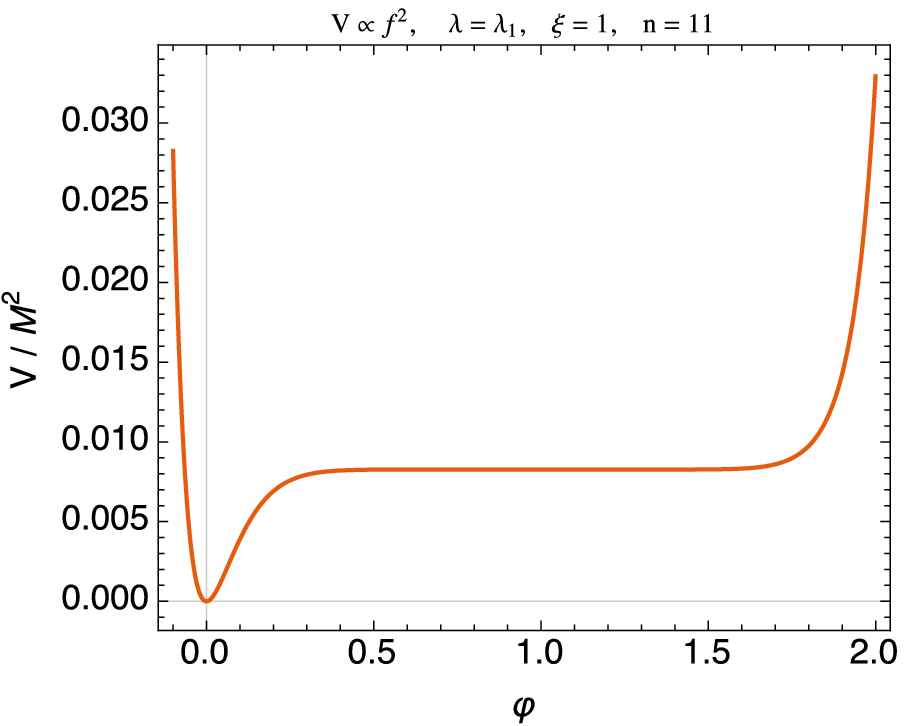} 
\hspace{0.5cm}
\includegraphics[height=4.6cm]{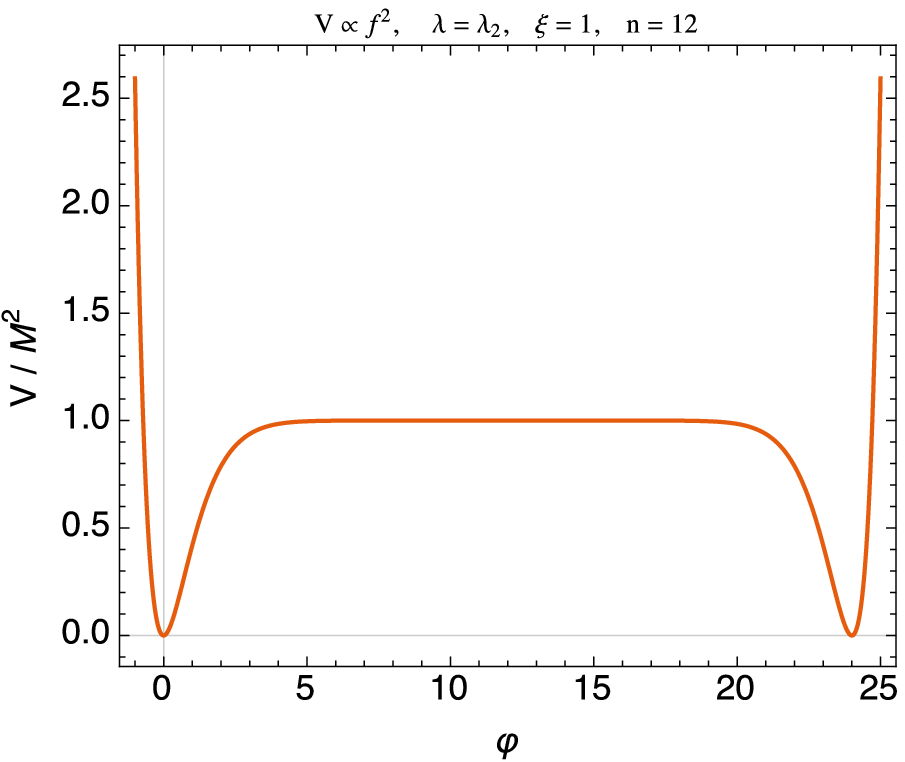}\\
\vspace{0.5cm}

\includegraphics[height=4.6cm]{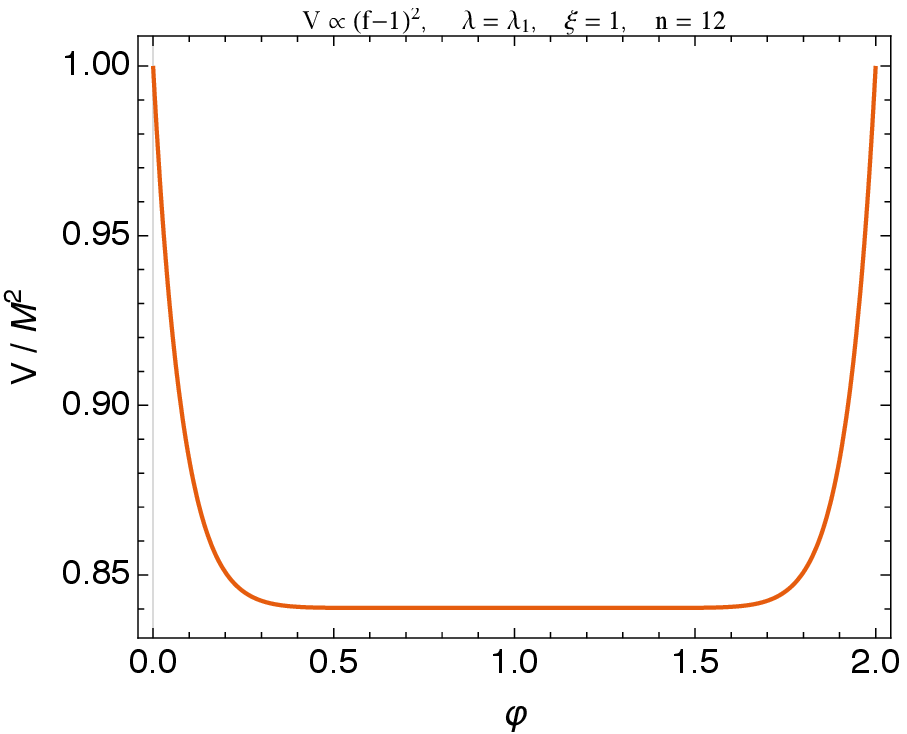} 
\hspace{0.5cm}
\includegraphics[height=4.6cm]{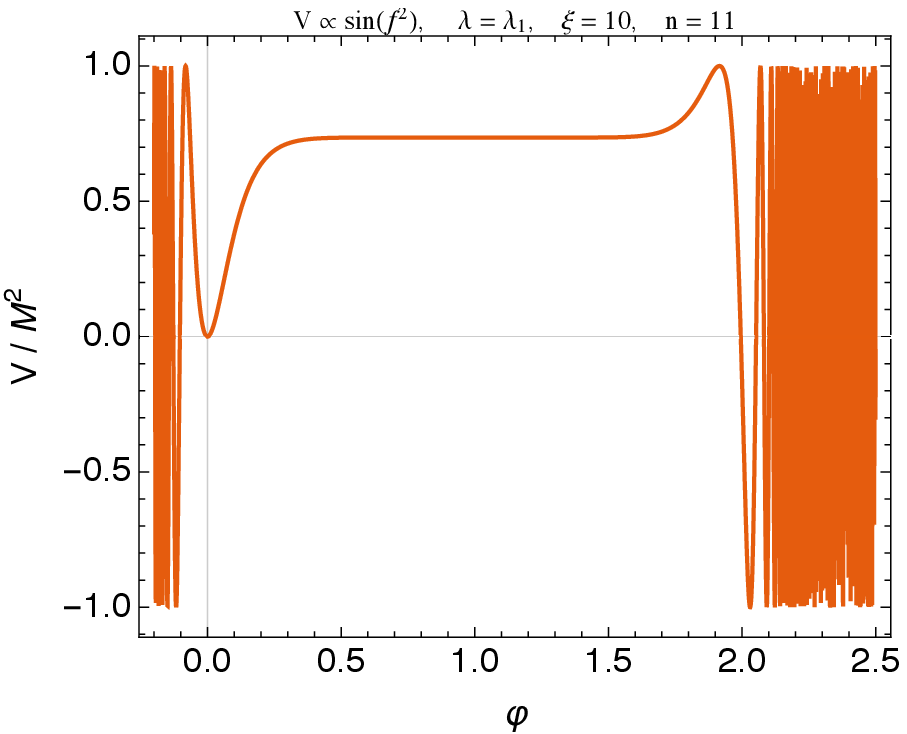}
\caption{\it The scalar potential $V = M^2 f^2$ for odd and even values of $n$ (upper left and upper 
right panels respectively), $V = M^2(f-1)^2$ and $V = M^2 \sin(f^2)$ (lower left and lower right 
panels respectively). The left panel shows that not every flat region can be used to generate inflation. For
even $n$ and $V \propto (f-1)^2$ one obtains a flat area of $V$ with positive vacuum energy density but 
without a graceful inflationary exit.} 
\label{fig:SaddleHilltop}
\end{figure}

 Some examples of the potentials that can constructed out of Eqs.~\eqref{eq:phin} and \eqref{eq:fn^-n} 
are shown in Fig.~\ref{fig:SaddleHilltop}. One can see that the stationary point of $f(\varphi)$ gives
rise to \textit{locally} flat regions able to support inflation. Note however this condition does not guarantee 
the existence of a graceful inflationary exit.

\begin{figure}
\centering
\includegraphics[height=4.6cm]{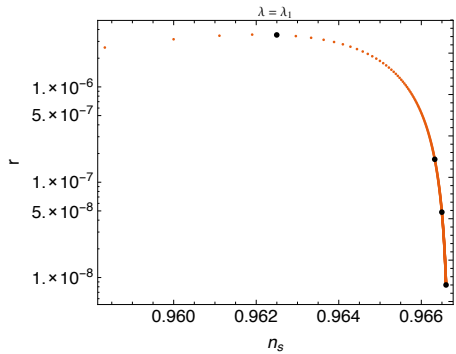} 
\caption{\it The tensor-to-scalar ratio $r$ and a spectral index $n_s$ for $\lambda = \lambda_1$ and 
different values of $n$. The result is $\xi$-independent. Black dotes denote $n=10$, $n=100$, $n=200$ 
and $n=500$ (from the top to the bottom). Taking big values on $n$ leads to big values of $\lambda_k$, which 
rises the question of perturbativity in the $\lambda = \lambda_1$ case.} 
\label{fig:Pert}
\end{figure}

To illustrate the influence of $n$ on the primordial inhomogeneities and on the perturbativity of the theory, consider 
for instance the inflationary observables associated to a potential $V = M^2f^2$ with $\lambda = \lambda_1$ (see Fig.~\ref{fig:Pert}). The amplitude of the primordial power-spectrum, the spectral tilt and the tensor-to-scalar ratio  read  
respectively
\begin{equation}
\Delta_\mathcal{R} = \frac{M^2 \xi ^2}{48 \pi ^2 n^4}\left(2(n-2)nN_\star\right)^{\frac{2 (n-1)}{n-2}}\, ,
\end{equation}

\begin{eqnarray}
n_s = 1-\frac{2 (n-1)}{(n-2) N_\star}  \, , \qquad r = 32 n^2 (2 (n-2) n N_\star)^{-\frac{2 (n-1)}{n-2}} \, ,
\end{eqnarray}
with $N_\star$ the number of e-folds at horizon crossing. Both $n_s$ and $r$ turn out to 
be independent of $\xi$ and $M$, but the scale of inflation (proportional to $r$) decreases with  $n$. A 
small inflationary scale would translate into large values of the effective coupling constants of the 
scalar potential ($\tilde{\lambda}_k = \frac{1}{k!}\frac{d^kV}{d \phi^k}$),
compromising with it the perturbativity of the theory.

The $n \to \infty$ limit of Eq.~(\ref{eq:fn^-n}) is 
particularly interesting. In this case, the deviations of $f(\varphi)$ from a constant value become 
exponentially suppressed
\begin{equation}
f(\varphi) = 1 - e^{-\xi \varphi} \, . \label{eq:ninf}
\end{equation}
This result holds even after multiplying $\lambda$ by any positive, $n$-independent constant. The 
resulting class of models is made of asymptotically flat potentials. A particular 
example is $V = M^2 f(\varphi)^2$, which gives rise to the Starobinsky-like 
potential
\begin{equation}\label{sta_pot}
V(\varphi) = M^2\left(1-e^{-\xi \varphi}\right)^2 \,.
\end{equation}
Note that the saddle point in Eqs.~\eqref{eq:ninf} and \eqref{sta_pot} appears at $\varphi \to \infty$. This behaviour 
coincides with the one obtained in Refs.~\cite{Artymowski:2015pna,Artymowski:2016mlh}, where the authors considered
the appearance of stationary points in $f(R) = \sum_{k=1}^{n} \alpha_k R^k$ modified gravity theories. As shown in this works, 
the inclusion of an infinity number of terms in the power series, together with the requirement 
of maximal flatness, translates into the appearance of a Starobinsky-like potential with exponentially 
suppressed corrections and into a displacement of the saddle point towards $R \to \infty$. In the large $n$ limit,  
both $f(R)$ and the scalar theory \eqref{eq:lagrang} give rise asymptotically flat Einstein frame potentials. Note 
however that the scalar sector generates a different Einstein frame potential than the one obtained by requiring
extreme flatness in  $f(R)$. In the $f(R)$ approach, it is not  possible to write the Ricci scalar as an analytical 
function of the scalaron field. The Einstein frame potential can be expressed only as 
a function of $R$.

\section{Equivalence to $\alpha$-attractors} \label{sec:alpha}

The supergravity embeddings considered in Ref.~\cite{Carrasco:2015rva} may lead to non-canonical kinetic terms for
the scalar field playing the role of the inflaton. The associated Einstein-frame Lagrangian density reads 
\begin{equation}
\frac{\cal L}{\sqrt{-g}}= \frac{1}{2}R + \frac{(\partial \psi)^2}{\left(1-\frac{\psi^2}{6\alpha^2}\right)^2} - 
V(\psi) \,, \label{eq:alphaattractor}
\end{equation}
with $\alpha$ a free parameter taking any positive value.
In order to obtain a canonically normalized kinetic term one can perform the field redefinition
\begin{equation}
\psi = \sqrt{6\alpha}\tanh\frac{\Psi}{\sqrt{6\alpha}} \, . \label{eq:alphakin}
\end{equation} 
As a function of the new field, the potential stretches around the pole in \eqref{eq:alphaattractor} 
and becomes \textit{locally} flat. As shown in Refs.~\cite{Galante:2014ifa,Rinaldi:2015yoa,Terada:2016nqg} and \cite{Kallosh:2013daa,Karananas:2016kyt}, this 
interesting property can be extended to higher order poles and to models containing more than one scalar field. 
The basic assumption 
about the potential is that it is must be a regular function of $\psi$ around the pole. Something 
similar happens in our case.  For seeing this explicitly, consider the field redefinition 
\begin{equation}
\varphi(f) = (\lambda \, n)^{\frac{1}{1-n}} \left(1-\left(1-\frac{f \, 
n (\lambda \, n)^{\frac{1}{n-1}}}{\xi }\right)^{1/n}\right) \, .
\end{equation}
This field transformation gives rise to the following kinetic term for the $f$ function, which plays now the role 
of the inflaton,
\begin{equation}\label{fkin}
(\partial \varphi)^2 = \frac{1}{\xi ^2}\left(\frac{\xi }{f \, 
n (\lambda \,  n)^{\frac{1}{n-1}}-\xi }\right)^{\frac{2 (n-1)}{n}} (\partial f)^2 \, .
\end{equation}
As in Eq.~\eqref{eq:alphaattractor}, the denominator of this expression contains a pole of order 
$2(n-1)/n$ which stretches the potential for the canonically normalized field $\varphi$. This stretching is 
localized in the vicinity of
\begin{equation}
f_p = \frac{\xi}{n}\left(\lambda \, n\right)^{-\frac{1}{n-1}} \, ,
\end{equation}
which is  precisely the value of $f$ at the stationary point $\varphi_s$, i.e. $f(\varphi_s)$, see Eq.~\eqref{eq:fgeneral}. The 
stretching of the potential around the pole is therefore equivalent to the existence of a stationary point in $f(\varphi)$. 
In both cases, the shape of the potential is irrelevant as long as it is regular and positive in the vicinity of the 
pole/stationary point. 

Exponentially flat potentials appear when the pole in Eq.~\eqref{fkin} is quadratic, i.e. in the 
$n\rightarrow\infty$ limit. In order to employ that limit, the function $f(\varphi)$ must be well defined, which 
is not guaranteed for a general $\lambda$. Among the cases \eqref{eq:phin} and \eqref{eq:fn^-n}, the particular choice
that provides us with a finite and regular $f(\varphi)$ at $n\rightarrow\infty$ is $\lambda = \lambda_2$. In this case,
Eq.~\eqref{fkin} becomes
\begin{equation}
(\partial\varphi)^2 = \frac{(\partial f)^2}{\xi^2(f-1)^2} \, ,
\end{equation}
which is { very similar to} the kinetic term of the $\alpha$-attractors \eqref{eq:alphaattractor}. To obtain 
the form appearing in this equation it is enough to perform an additional field redefinition
\begin{equation}
\psi(f) = \frac{\sqrt{6 \alpha } \left((6 \alpha  (1-f) \xi )^{\sqrt{\frac{2}{3 \alpha }} \xi }-1\right)}{(6 \alpha  (1-f) \xi )^{\sqrt{\frac{2}{3 \alpha }} \xi }+1} \, .
\end{equation}
Since we have not assumed much about $V(f)$, we can still keep it as a general function of $\psi$ after this change of variables.
Note that the equivalence between the appearance of poles in kinetic terms and the flatness of the potential in terms 
of a canonically normalized field has been also analysed in Ref. \cite{Broy:2015qna}. Nevertheless, the approach 
presented here is significantly different. In particular, we took a model building perspective based on the existence 
of a multi-stationary point able to ensure the flatness of the potential rather than invoking a particular pole structure
on the kinetic terms.

%%%%%%%%%%%%%%%%%%%%%%%%%%%%%%%%%%%%%%%%%%%%%%%%%%
%%%%%%%%%%%%%%%%%%%%%%%%%%%%%%%%%%%%%%%%%%%%%%%%%%

\section{Summary} \label{sec:Summary}

In this paper we considered a general scalar potential $V(f(\varphi))$ with $f(\varphi) = \xi\sum_{k=1}^n \lambda_k\varphi^k$. 
By requiring the existence of the $m$-order stationary point at a field value $\varphi = \varphi_s$, we obtained a 
specific form for $f(\varphi)$ containing three free parameters $\lambda\equiv\lambda_n$, $\xi$ and $n$. We showed 
that around the stationary point $\varphi_s$, the inflationary potential is flat and suitable for inflation 
provided that $V(\varphi_s) > 0$. Nevertheless, not all of the resulting flat potentials allow for a
graceful inflationary exit. 

The relation between our results and the theory of $\alpha$-attractors was also considered. We explicitly showed that 
using $f$ as a scalar field it is possible to obtain a non-canonical kinetic term, which after a trivial field 
redefinition, takes the form of the $\alpha$-attractor kinetic term. There is therefore 
a deep connection between the two approaches: asking for the existence of a stationary point in a scalar potential
with canonical kinetic term is equivalent to ask for the existence of a kinetic term with a pole. Both formulations
leads to flat regions within general potentials which can be responsible for inflation.

%%%%%%%%%%%%%%%%%%%%%%%%%%%%%%%%%%%%%%%%%%%%%%%%%%
%%%%%%%%%%%%%%%%%%%%%%%%%%%%%%%%%%%%%%%%%%%%%%%%%%

\section*{Acknowledgements}

MA was supported by National Science Centre grant FUGA UMO-2014/12/S/ST2/00243. JR acknowledges
support from DFG through the project TRR33 ``The Dark Universe'' and thanks Georgios Karananas for 
useful comments on the manuscript.

\end{document}